\documentclass[fleqn,usenatbib]{mnras}

\usepackage{newtxtext,newtxmath}

\usepackage[T1]{fontenc}

\DeclareRobustCommand{\VAN}[3]{#2}
\let\VANthebibliography\thebibliography
\def\thebibliography{\DeclareRobustCommand{\VAN}[3]{##3}\VANthebibliography}

\usepackage{graphicx}	
\usepackage{amsmath}	

\defcitealias{son_strong_2025}{S25}
\defcitealias{Wiseman2022}{W22}
\usepackage{lineno}

\title[SN Ia cosmology robust to age evolution]{Still Accelerating: Type Ia supernova cosmology is robust to host galaxy age evolution}

\author[P. Wiseman et al.]{
\parbox{\textwidth}{
\Large
Phil Wiseman$^{1}$\thanks{E-mail: p.s.wiseman@soton.ac.uk (PW)},
Brodie Popovic$^{1}$,
Mark Sullivan$^{1}$,
Adam G.~Riess$^{2,3}$,
Dan Scolnic$^{4}$,
Rebecca C.~Chen$^{5,6,7}$,
Tamara M.~Davis$^8$,
Llu\'is Galbany$^{9,10}$,
Isobel M.~Hook$^{11}$,
Saurabh W. Jha$^{12}$,
Lisa Kelsey$^{13}$,
Yukei S.~Murakami$^{3}$, 
Micka\"el Rigault$^{14}$,
Benjamin M.~Rose$^{15}$,
Brian Schmidt$^{16}$,
Mat Smith$^{11}$,
Maria Vincenzi$^{17}$
}
\vspace{0.04in}\\
$^{1}$School of Physics and Astronomy, University of Southampton, Southampton SO17 1BJ, UK\\
$^{2}$Space Telescope Science Institute, 3700 San Martin Drive, Baltimore, MD 21218, USA\\
$^{3}$Department of Physics and Astronomy, Johns Hopkins University, Baltimore, MD 21218, USA\\
$^{4}$Department of Physics, Duke University, Durham, NC 27708, USA\\
$^{5}$Kavli Institute for Particle Astrophysics \& Cosmology, Stanford University, Stanford, CA 94305, USA\\
$^{6}$SLAC National Accelerator Laboratory, Menlo Park, CA 94025, USA\\
$^{7}$Brinson Prize Fellow\\
$^8$School of Mathematics and Physics, University of Queensland, Brisbane, QLD 4072,  Australia\\
$^{9}$Institute of Space Sciences (ICE, CSIC), Campus UAB, 08193 Barcelona, Spain\\
$^{10}$Institut d’Estudis Espacials de Catalunya (IEEC), 08034 Barcelona, Spain\\
$^{11}$Department of Physics, Lancaster University, Lancaster LA1 4YB, UK\\
$^{12}$Department of Physics and Astronomy, Rutgers, the State University of New Jersey, Piscataway, NJ 08854, USA\\
$^{13}$Institute of Astronomy and Kavli Institute for Cosmology, University of Cambridge, Cambridge CB3 0HA, UK\\
$^{14}$Universit\'e Claude Bernard Lyon 1, CNRS, IP2I Lyon / IN2P3, IMR5822, F-69622 Villeurbanne, France\\
$^{15}$Department of Physics and Astronomy, Baylor University, Waco, TX 76798, USA\\
$^{16}$Research School of Astronomy and Astrophysics, Australian National University, Canberra, ACT 0200, Australia\\
$^{17}$Astrophysics, Department of Physics, University of Oxford, Keble Road, Oxford OX1 3RH, UK\\
}

\date{Accepted XXX. Received YYY; in original form ZZZ}

\pubyear{\the\year{}}

\begin{document}
\label{firstpage}
\pagerange{\pageref{firstpage}--\pageref{lastpage}}
\maketitle

\begin{abstract}
Type Ia supernovae are a cornerstone of modern cosmology, providing first evidence for cosmic acceleration and new tests of dark energy.  Son et al. 2025 (S25) claim a strong redshift evolution in standardized supernova luminosities driven by supernova progenitor age, with dramatic cosmological implications: rapidly evolving dark energy, decelerating expansion, and a $9\sigma$ tension with $\Lambda$CDM. We show that the underpinning evidence required for this conclusion -- the supernova progenitor-age dependence, the redshift-dependent age difference, and their combined impact -- is either negligible or relies on effects already corrected for in modern supernova analyses. First, the S25 analysis omits the standard host-galaxy stellar mass correction that captures known environmental dependencies that also correlate with stellar age. Applying this correction to the S25 sample, we find no dependence of standardized supernova brightness on host age. Independent data also show no significant difference at low-redshift in standardized brightness between star-forming galaxies and several Gyr older quiescent galaxies of the same stellar mass. Second, the S25 scenario predicts strong redshift evolution of the host-mass effect. Data from the Dark Energy Survey supernova survey measure evolution of $-0.028 \pm 0.034~\mathrm{mag}\,z^{-1}$, consistent with zero and altering the dark-energy equation-of-state measurement ($w$) by $<$0.01 if included. Third, we demonstrate that the claimed $\sim5$~Gyr progenitor age difference between nearby and distant supernovae is overstated by factors of three to five largely due to a conflation of host galaxy age with supernova progenitor age. We conclude that type~Ia supernova cosmology remains robust for current measurements of dark energy.

\end{abstract}



\section{Introduction}

Type Ia supernovae (SNe Ia) are one of the best standardizable candles and most powerful probes of cosmology. Measurements of SN Ia distances and redshifts provide direct observational evidence of cosmic acceleration
and the inference of dark energy \citep{Riess1998, Perlmutter1999}.  Using SNe Ia alone, the evidence for acceleration is now very strong, and precise constraints on the dark energy equation-of-state parameter $w$ are routinely obtained using SNe Ia in combination with complementary probes \citep{Brout2022, des_collaboration_dark_2024,rubin_union_2025, popovic_dark_2025}, while evidence for dark energy also remains strong without using SNe \citep{desi_collaboration_desi_2025, des_collaboration_dark_2025}. SNe also remain an important component in measurements of the Hubble constant \citep{riess_comprehensive_2022,galbany_updated_2023,breuval_small_2024,freedman_status_2025,h0dn_collaboration_local_2025}.

The use of SNe Ia as distance indicators relies, broadly, on the ability to standardize their luminosities and that the standardized luminosity does not change significantly with cosmic time (i.e., redshift).  The standardization itself typically exploits two empirical relations: the `faster--fainter' relation between the SN Ia peak brightness and the width of its light curve \citep{Phillips1993}, and the \lq bluer--brighter\rq\ relationship between the SN Ia peak brightness and its colour \citep{Riess1996, Tripp1998}. With modern datasets and analysis techniques, the post-standardization scatter on the inferred distances to SNe Ia unexplained by model or measurement uncertainties is $\sim 0.1$\,mag in distance modulus, or $\sim$5 per cent in distance.

This standardization process leaves a residual dependence of the SN Ia distances on the properties of the galaxies in which they exploded: the post-standardization peak brightness of SNe Ia in \lq low stellar mass\rq\ ($\log(M_{\star}/\mathrm{M}_{\sun}) \leq 10$) galaxies is, on average, $\sim 0.05-0.1$ mag fainter than in their higher stellar mass counterparts regardless of survey, redshift or analysis approach \citep{Kelly2010, Sullivan2010,Lampeitl2010,childress_host_2013,uddin_influence_2017,ramaiya_dependence_2025}. As a result, the inclusion of observed SN--host relationships in the estimation of SN distances, in some form, is now well-established in SN Ia cosmology \citep{Conley2011,Betoule2014,Brout2022,vincenzi_dark_2024, rubin_union_2025}.

It is widely acknowledged that stellar mass is unlikely to be the driving cause of this effect, but the actual underpinning physical cause remains unclear. Empirically, there are well-known relationships between galaxy stellar mass and many other galaxy properties \citep[e.g.,][]{Tremonti2004,Gallazzi2005,mannucci_fundamental_2010,garn_predicting_2010} making the isolation of the causal variable difficult. Dependencies between standardized SN Ia luminosity and other host galaxy properties are observed, including gas-phase metallicity \citep{Gallagher2005, dandrea_spectroscopic_2011,childress_host_2013, pan_host_2014,Moreno-Raya2016,Millan-Irigoyen2022}, stellar metallicity \citep{Gallagher2008}, star-formation activity \citep[][]{Sullivan2010, Rigault2013,Rigault2018}, rest-frame galaxy colour \citep{childress_host_2013,Roman2018,Kelsey2021,kelsey_concerning_2023,ginolin_ztf_2025}, and the equivalent width of the [\ion{O}{ii}] emission line \citep{Dixon2022,martin_o_2024}. 

Many of these observations have indirect links to stellar age, and this has motivated significant research on the link between SN Ia progenitor ages, host galaxy stellar ages, and SN Ia standardization \citep{Rigault2013,childress_host_2013,Rose2019,Wiseman2022,Wiseman2023}. There are two signficant challenges. The first is that galaxy ages are notoriously difficult to measure without very high-S/N galaxy spectroscopy, and fits to broad-band optical photometry only estimate luminosity-weighted ages of the stellar population. This is only the same as the SN progenitor age in a coeval simple stellar population, i.e., a group of stars formed at the same time; galaxies universally have more complex star-formation histories (SFHs). The second challenge is the function describing the probability of a stellar population producing a SN Ia as a function of time: the delay-time distribution \citep[DTD; see review of][]{Maoz2014}. The rate of supernovae in a galaxy is a convolution of the SN Ia DTD and the galaxy's SFH, making the interpretation of galaxy ages in the context of SN progenitors extremely complex: a typical massive galaxy can produce SNe Ia from both young and old progenitor systems. Reflecting this uncertainty, SN cosmology analyses  apply standardization corrections based on stellar mass as a simple empirical variable, and SN progenitor and galaxy age models can then be included as systematic tests; see, for example, the Dark Energy Survey (DES) 5-year SN cosmology analysis \citep[DES-SN5YR;][]{vincenzi_dark_2024,popovic_dark_2025}.


However, a series of studies by the same group \citep{kang_early-type_2020,lee_evidence_2022,chung_root_2023,chung_strong_2025}, have claimed that there is a linear trend whereby SNe Ia in older stellar populations (older host galaxies) are brighter post-standardization compared to those in younger host galaxies, and more critically that this trend is unaccounted for in cosmological analyses. Since stellar populations evolve with redshift, if this correlation is intrinsic to the SNe then it could induce an apparent redshift evolution in the post-standardization SN Ia brightnesses, biasing cosmological measurements \citep[see, e.g.,][]{sarkar_implications_2008,Rigault2013,Childress2014,Rigault2018}. \citet{lee_evidence_2022} argued that applying a redshift-dependent correction for this effect could eliminate the evidence for dark energy. More recently, \citet[][hereafter S25]{son_strong_2025} report a strongly-evolving dark energy based on a correction derived from data from \cite{chung_strong_2025}, with the dark energy equation-of-state parameter $w$ showing significant redshift evolution.

These claims are important and warrant a thorough examination and testing. This serves as the motivation for this paper and we examine the claims in \citetalias{son_strong_2025} in two contexts. Firstly, in Section \ref{sec:state_of_the_art} we describe the established tools that are used in  cosmological SN Ia analyses and that are designed to mitigate for host-galaxy biases. Since galaxy stellar mass and galaxy age are highly correlated, it is not surprising, and we show, that the mass-standardization already included in modern SN Ia analyses largely accounts for all SN--host correlations, leaving no substantial leftover correlation with host galaxy ages. Secondly, in Section \ref{sec:age_review} we review the complex relationships between SN Ia progenitors and their host galaxies and show several subtle but critical issues with the method involved in applying a redshift-dependent correction to SN Ia luminosities based on the host galaxy age measurements. We summarize with a conclusion in Section \ref{sec:conclusion}.


\begin{figure*}
    \centering
    \includegraphics[width=0.32\textwidth]{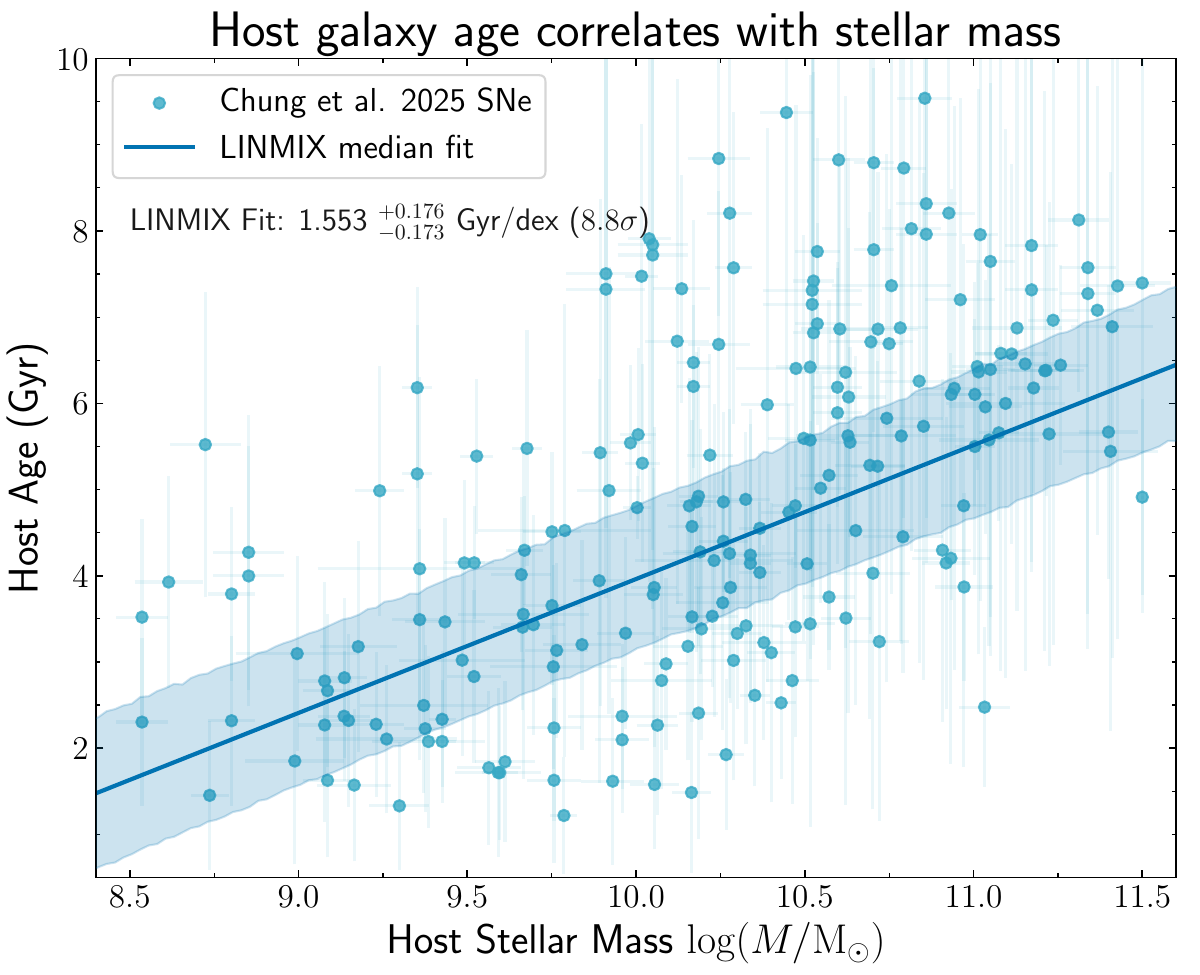}
    \includegraphics[width=0.32\textwidth]{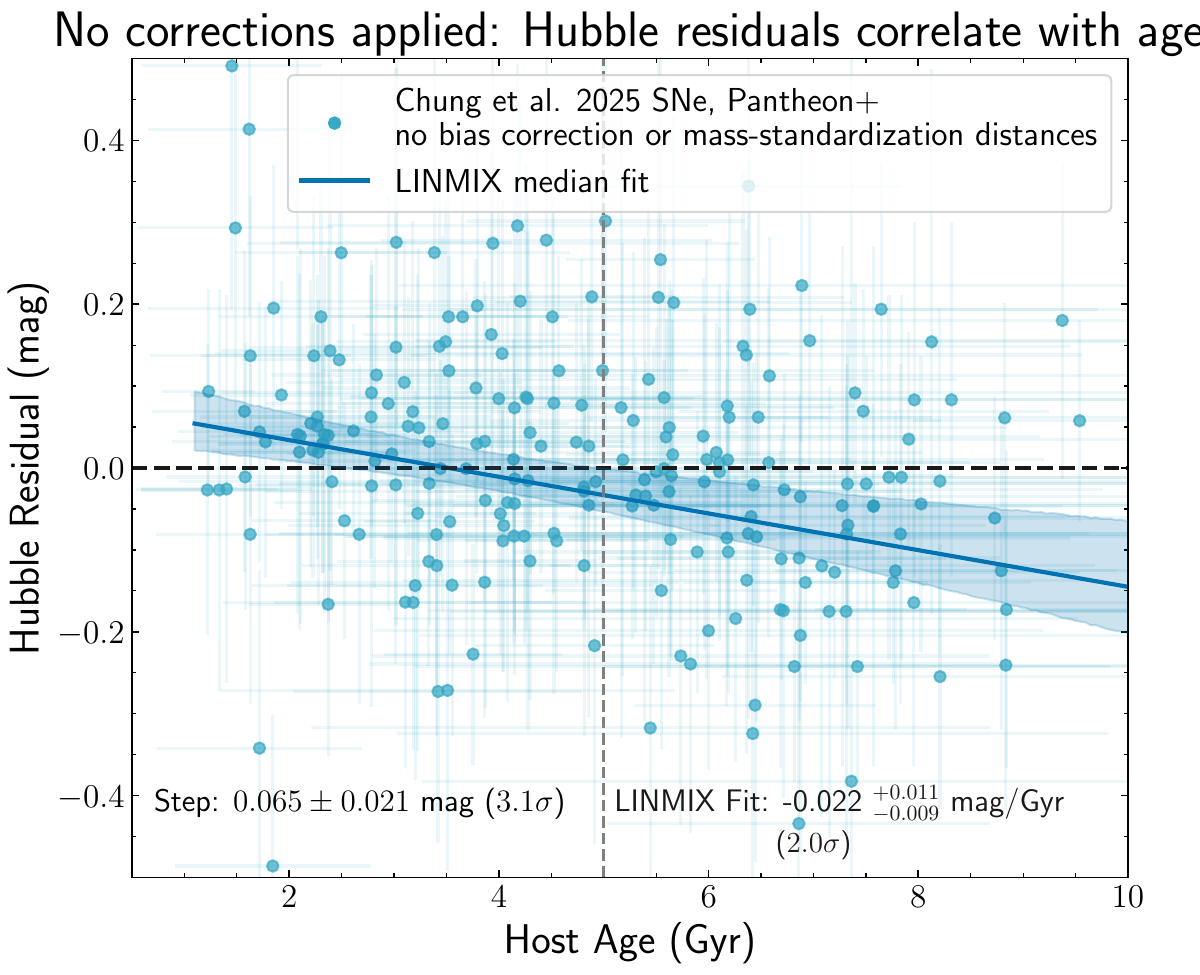}\includegraphics[width=0.32\textwidth]{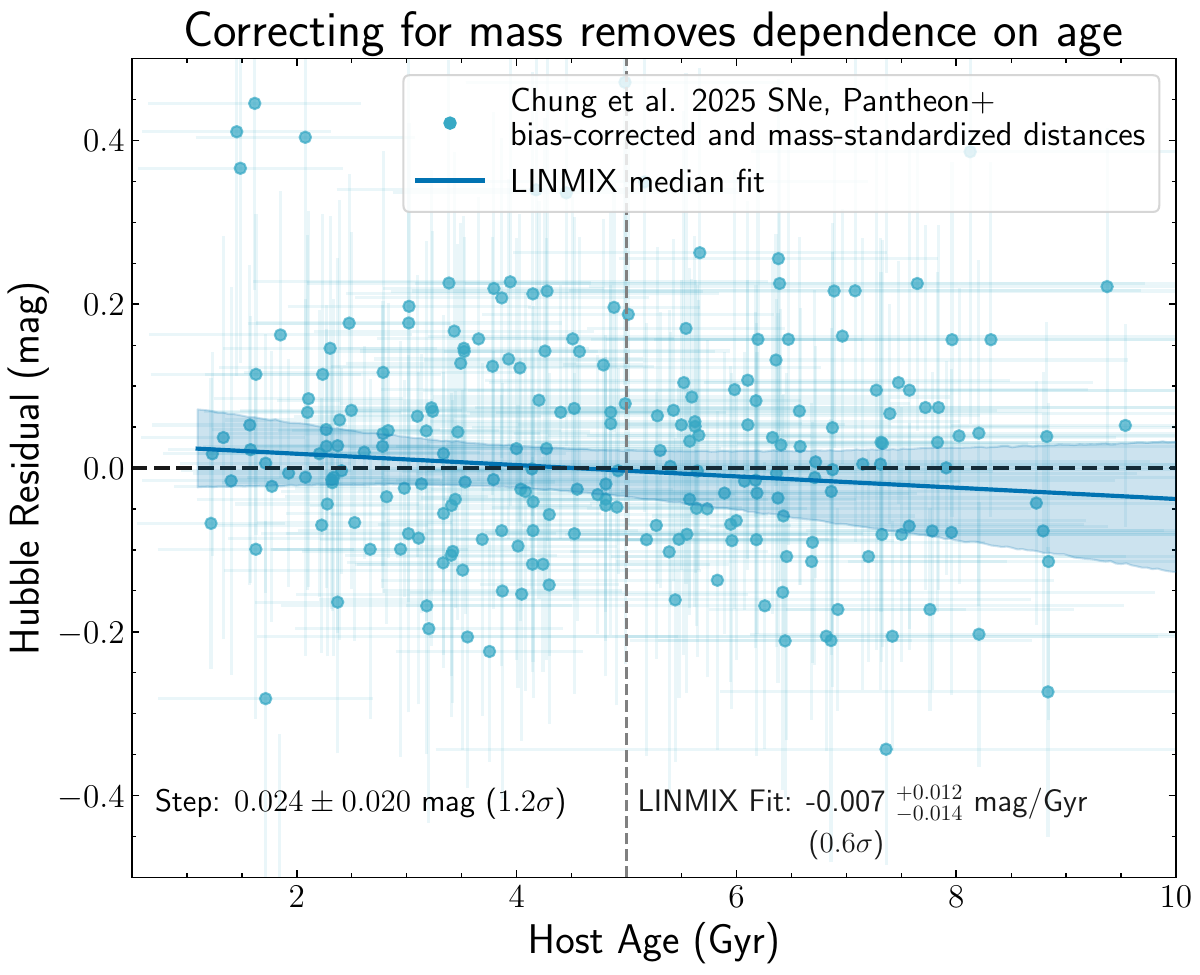}

\caption{\textit{Left: }Host galaxy stellar mass versus galaxy age for the sample used by \citetalias{son_strong_2025}. Galaxy ages are taken from \citet{chung_strong_2025} and are strongly correlated with host galaxy stellar mass. {\it Centre:} Hubble residual before bias correction and mass standardization, versus galaxy age. Hubble residuals are taken from the Pantheon+ analysis. We recover a somewhat smaller slope than \citet{chung_strong_2025}, and with lower significance;
{\it Right:} Hubble residuals after a bias correction and standardization for stellar mass. The relationship between Hubble residual and host galaxy age is smaller, and not significant. 
\label{fig:ages_and HRs}}

\end{figure*}

\section{Robustness of SN Ia cosmology to host galaxy correlations}
\label{sec:state_of_the_art}

In this section we review the procedures underpinning SN Ia cosmology, and highlight how modern cosmology analyses with SNe Ia detect and mitigate the host-galaxy driven biases discussed in \citetalias{son_strong_2025}. We show that these analyses find no significant evidence for redshift evolution in SN Ia standardization.

\subsection{SN Ia distance measurements}

SNe Ia can be standardized by applying adjustments based on the `faster--fainter' and `bluer--brighter' relationships to their measured peak brightnesses $m_B$. There are different frameworks to implement this. In this paper we infer distance moduli, $\mu_{\rm obs}$, via an adapted version of the relation presented by \citet{Tripp1998}, i.e.,
\begin{equation}
    \mu_{{\rm obs},i} = m_{B,i} - M_0 + \alpha x_{1,i} -\beta c_i + \gamma G_{{\rm host},i} +\mu_{{\rm bias},i} \,,
\label{eq:simple_tripp}
\end{equation}
where $x_{1,i}$ and $c_i$ are the light curve \lq stretch\rq\ and colour\footnote{Here we present the Tripp formula in the framework of fitting SNe Ia with the Supernova Adaptive Light Curve Template (SALT2; \citealt{Guy2007,Guy2010,kenworthy_salt3_2021}), but the principle is the same regardless of the fitting technique and exact definitions of the SN parameters.} of the $i$th SN, $\alpha, \beta$ parametrize the global standardization relationships, and $M_0$ is the global intrinsic SN Ia absolute magnitude for $x_1=0$ and $c=0$. $G_{{\rm host},i}$ represents some property $P$ of the $i$th SN's host galaxy, often stellar mass, and $\gamma$ represents the size of the correction based on that host property, typically applied as a step function at some threshold $P_\mathrm{step}$, i.e.,
\begin{equation}
\gamma G_\mathrm{host}=
\begin{cases}
+\gamma/2&\mathrm{if}~G_\mathrm{host}>P_\mathrm{step},\\
-\gamma/2&\mathrm{otherwise},
\end{cases}
\label{eq:Ghost}
\end{equation}
with the correction often colloquially referred to as the \lq mass step\rq. The $\mu_{{\rm bias},i}$ term is a bias correction made to each SN to correct for both Malmquist-like selection effects \citep[e.g.,][]{hamuy_selection_1999,Marriner2011} and astrophysical effects that include relationships between SNe Ia and their hosts and dust extinction \citep{Brout2020,Popovic2021}, and is calculated from simulations \citep{Kessler2009,Kessler2017}.

The \lq Hubble residuals\rq, $\Delta\mu$, are then defined as the difference between $\mu_{\rm obs}$ and the distance modulus calculated from a cosmological model, $\mu_\mathrm{theory}$, defined as
\begin{equation}
    \Delta\mu=\mu_\mathrm{obs}-\mu_\mathrm{theory}\left(\mathcal{C},z\right)
\end{equation}
where $\mathcal{C}$ is a set of cosmological parameters. The dispersion or scatter of $\Delta\mu$ around the best cosmological fit is caused by a combination of observational noise and a remaining inherent dispersion of the SNe, often termed `intrinsic scatter'. The magnitude of intrinsic scatter and its relationship to SN and their host galaxy parameters is well studied and accounted for in cosmological measurements \citep[e.g.,][]{Guy2010, Chotard2011,Brout2020,Popovic2021a}.

\subsection{Overview of the S25 argument}

\citetalias{son_strong_2025} assemble a heterogeneous sample of SNe Ia, taking SNe and their host galaxy photometry from two samples of SNe discovered by the Sloan Digital Sky Survey (SDSS) SN Survey: \citet{gupta_improved_2011} and \citet{Rose2019}. The Hubble residuals of these SNe were compiled for an analysis by \citet{chung_strong_2025}, that in turn were originally published by \citet{gupta_improved_2011} and \citet{campbell_cosmology_2013}, the latter using an early implementation of SN Ia photometric classification methods. Neither of those studies included any host galaxy (stellar mass) correction term or the bias correction term in equation ~\ref{eq:simple_tripp}. Instead, \citet{chung_strong_2025} use a single ad hoc corrective term as a function of redshift in an attempt to account for an expected astrophysical bias from progenitor-redshift evolution. The Hubble residuals are then compared to estimates of host-galaxy stellar population ages derived from spectral energy distribution (SED) template fitting to broad-band optical (SDSS) host galaxy photometry \citep{chung_strong_2025}.

\citetalias{son_strong_2025} show that their Hubble residuals correlate with the estimated ages of the host galaxies. \citetalias{son_strong_2025} then argue that evolution in the mean age of the host population with redshift will result in incorrect distance measurements for the SNe, and biased cosmological parameters. \citetalias{son_strong_2025} estimate the magnitude of this bias by assuming that the Hubble residual -- galaxy age relation translates directly to a Hubble residual -- SN Ia progenitor age relation, and then derive a \lq correction\rq\ based on a model of SN Ia progenitor age redshift evolution. This methodology is inaccurate, as the SN progenitor age will always be younger than the age of its host galaxy, and the relation between them will itself vary with redshift and depends on modelling assumptions. We return to this topic in Section~\ref{sec:age_review}.

\begin{figure}
    \centering
    \includegraphics[width=0.48\textwidth]{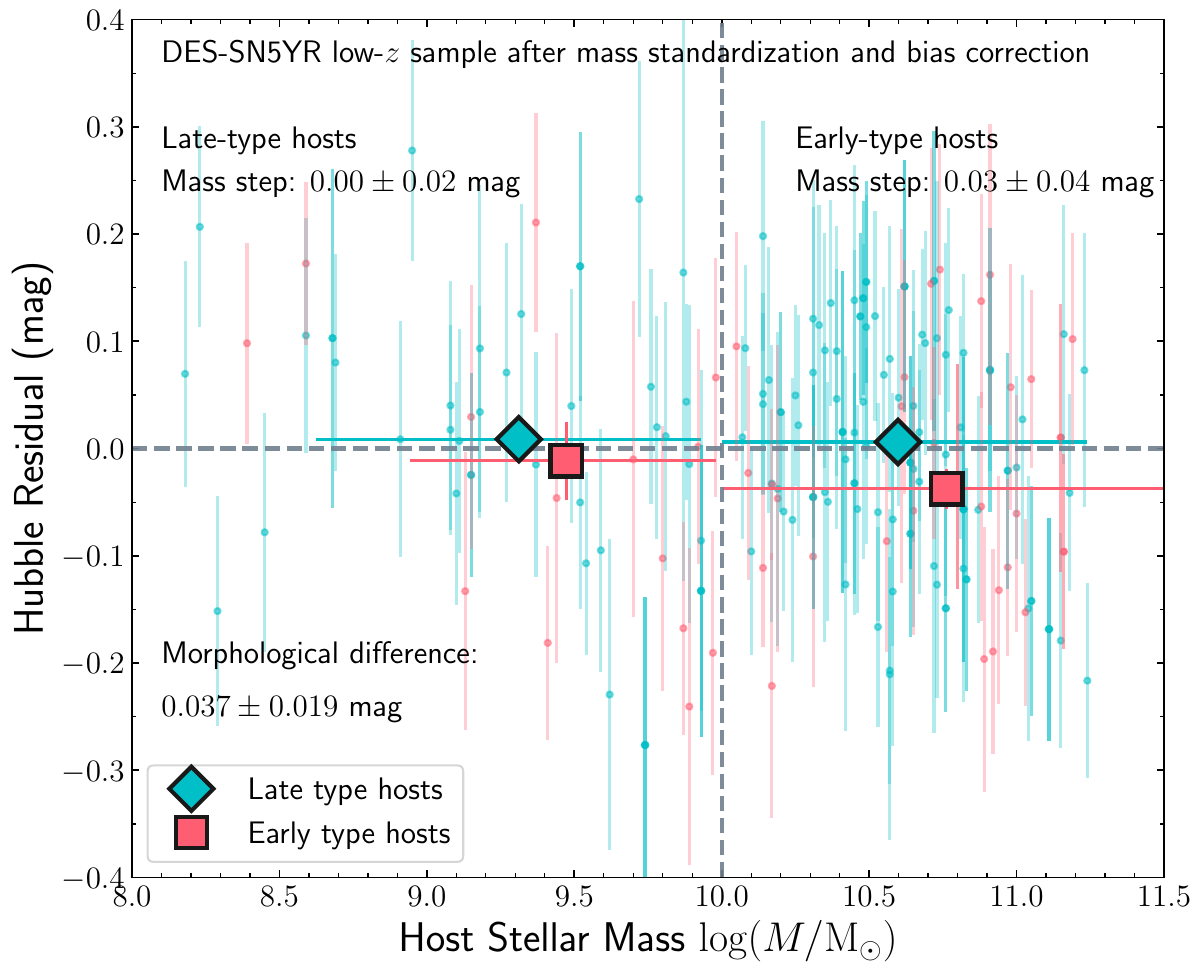}

\caption{Hubble residuals of low-redshift SNe Ia as a function of stellar mass, split by the morphology of the host galaxies. Hubble residuals are from the DES-SN5YR compilation and have had both the stellar mass standardization and bias correction applied 
The mean difference in age of the two morphology categories is several Gyr, which based on the \citetalias{son_strong_2025} scenario of 0.03\,mag\,Gyr$^{-1}$ would predict a $\sim$0.09--0.18\,mag difference between the two populations, which is not seen in the mean of the data.
\label{fig:HRs_vs_mass}}
\end{figure}



\subsection{Testing the existing framework for host galaxy -- SN standardization relationships}
\label{subsec:existing_corrections}

In modern SN Ia analyses distance estimates account for host galaxy dependencies in two ways, both neglected by \citetalias{son_strong_2025}. The first is the straightforward application of the $G_\mathrm{host}$ term in equation~\ref{eq:simple_tripp}. The second is the modelling of selection biases ($\mu_\mathrm{bias}$ in equation~\ref{eq:simple_tripp}) in different SN samples, which are calculated from the relationships between SN properties (stretch, colour and brightness), dust extinction, host galaxies and survey characteristics.
Most SN Ia cosmology analyses use the host galaxy stellar mass as $G_\mathrm{host}$, primarily for its reliability and simplicity of measurement with the limited host galaxy photometric coverage typically available at high redshift \citep{bell_stellar_2001,taylor_masses_2010,salim_galex-sdss-wise_2016,ramaiya_dependence_2025}. We discuss two tests of this existing standardization framework.

\subsubsection{Tests using stellar mass as the standardization parameter}

Due to the nature of galaxy growth, there is a long-established, strong correlation between the stellar mass of a galaxy and the mean age of its stellar population: more massive galaxies are, on average, older. This relationship holds for SN Ia host galaxies and is shown in the left panel of Fig.~\ref{fig:ages_and HRs} for the SN sample used by \citet{chung_strong_2025}, where we find a correlation at $8.8\sigma$ significance using the \textsc{LINMIX} fitting method \citep{kelly_aspects_2007}.

While stellar age and stellar mass are clearly distinct host galaxy properties, the mass standardization and bias-correction terms remove any significant relation between Hubble residuals and host galaxy age for the SN sample used by \citetalias{son_strong_2025}.  For this sample we compare the galaxy ages as measured by \citet{chung_strong_2025} against more recent Hubble residuals published in the Pantheon+ SN Ia compilation for the same SNe \citep{Scolnic2021} and which, crucially, provides host (stellar mass) standardization and bias corrections for each event \citep{Brout2022}\footnote{\url{https://github.com/PantheonPlusSH0ES/DataRelease}, commit 7fc6805}. We use the Pantheon+ data as it contains Hubble residuals for the SDSS SNe used by \citet{chung_strong_2025}, unlike the more recent DES-SN5YR compilation.

To show the effect of these corrections, in Fig.~\ref{fig:ages_and HRs} we show the $\Delta\mu$--host age relationship for the \citet{chung_strong_2025} samples both before (centre) and after (right) stellar mass standardization and bias correction. Before correction, we find a slope $\sim$$25$ per cent smaller than that used in \citetalias{son_strong_2025} ($-0.022$\,mag\,Gyr$^{-1}$ compared to $-0.030$). After corrections, the slope $-0.007^{+0.012}_{-0.014}$\,mag\,Gyr$^{-1}$ is more than four times weaker than that measured by \citetalias{son_strong_2025} and is not statistically significant ($<1$$\sigma$).

Similar results have been found for the DES-SN5YR sample \citep{kelsey_concerning_2023} who use host galaxy rest-frame colour, a proxy for host age. They also find a small and not significant relationship between $\Delta\mu$ and host galaxy colour after correcting the $\Delta\mu$ for stellar mass, as well as vice-versa.


\subsubsection{Tests using SNe with different host galaxy morphologies}

A second simple test for a significant age bias in published SN Ia distances can be conducted by comparing the $\Delta\mu$ for SNe Ia in passive and star-forming galaxies in the local Universe. Galaxy morphology is not a perfect tracer of galaxy age \citep{Briday2021}, but early-type galaxies are known to contain older stellar populations than later-types at all stellar masses \citep{parikh_sdss-iv_2021}. In the model of \citetalias{son_strong_2025} the difference in the mean age of their stellar populations ($\sim$3--6\,Gyr) should be reflected in an age difference between SN progenitors: their models imply a $\sim$0.09--0.18\,mag difference in average $\Delta\mu$. Indeed, the lack of a difference in $\Delta\mu$ for local SNe Ia in early (old) and late (young) type hosts was one of the tests for evolution employed (and passed) in both \citet{Riess1998} and \citet{Perlmutter1999} for the initial evidence for cosmic acceleration: Riess et al., for example, measured a mean difference of $0.04\pm0.07$\,mag, smaller than expected by \citetalias{son_strong_2025}.

We use the DES-SN5YR SN sample \citep{sanchez_dark_2024}\footnote{from the main branch of \url{https://github.com/des-science/DES-SN5YR} using commit c9a4fca.} over $z=0.025-0.15$ and those SNe with host morphology visually classified \citep{riess_comprehensive_2022} and passing standard cosmological quality cuts \citep[e.g.,][]{Betoule2014}. In Fig.~\ref{fig:HRs_vs_mass} we show the Hubble residuals, after applying mass standardization and bias corrections, as a function of host stellar mass and split by morphological classification. On average, the $\Delta\mu$ difference between late-type hosted SNe ($N=149$ events) and early-type hosted SNe ($N=48$ events) is $0.037\pm0.020$\,mag, in the sense that SNe Ia hosted by early-type galaxies are brighter. The difference is almost identical ($0.036\pm0.021$\,mag) without applying mass standardization or bias correction. For a realistic $\sim$3--6 Gyr difference in mean age for these host types \citep{parikh_sdss-iv_2021, mattolini_re-assessing_2025} and the 0.03\,mag\,Gyr$^{-1}$ host-age -- $\Delta\mu$ gradient from \citetalias{son_strong_2025}, the difference is predicted to be $\sim$0.09--0.18\,mag. The DES-SN5YR data do not show evidence for the host galaxy age dependence at the level used by \citetalias{son_strong_2025} to infer, from those same data, their cosmological result.


These two tests demonstrate that the existing SN Ia cosmological framework appears to adequately correct for the claimed effects by \citetalias{son_strong_2025}, simply by accounting for known correlations between SNe Ia and their host galaxy stellar mass.

\subsection{Robustness to an evolving host galaxy correction} \label{subsec:gamma_z}

\begin{figure}
    \centering
    \includegraphics[width=0.48\textwidth]{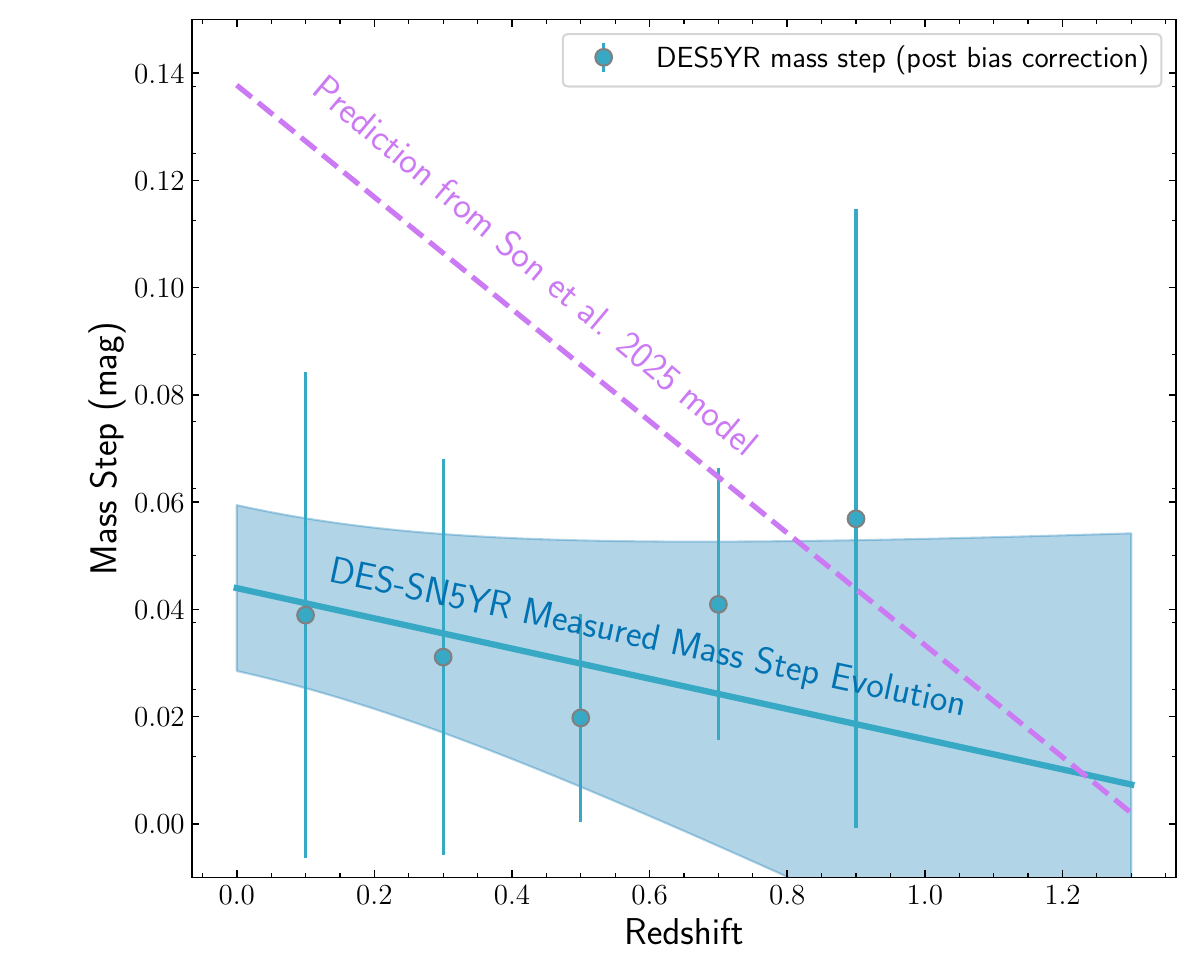}

\caption{The size of the redshift-evolving $\gamma$ (\lq mass step\rq) correction measured with DES-SN5YR data (blue points). The blue line is the best fit to the DES data of the form $\gamma(z)=\gamma_0 + \gamma_1 z$ and the blue shaded area the range of evolution allowed based on the uncertainty from this fit. This is compared to the predicted redshift evolution from the age-bias model of \citetalias{son_strong_2025} (purple line).  The important difference occurs at low-redshift where the observed $\gamma$ is much smaller than the prediction and little changed from high redshift.
\label{fig:gamma1}}
\end{figure}

We can further search for evidence of the failure of the current approach used in SN Ia cosmology by measuring the size of $\gamma$ in equation~\ref{eq:simple_tripp} in bins of redshift. This is an established test when inferring cosmological constraints from SN Ia datasets \citep[e.g.,][]{Betoule2014,rubin_union_2025}, and here we show data from the recent DES-SN5YR analyses \citep[][]{vincenzi_dark_2024,popovic_dark_2025} (Fig.~\ref{fig:gamma1}). If the primary driver of the relationship between host galaxy environment and SN standardization is the age of the SN progenitor or of the galaxy, then the size of $\gamma$ will decrease with increasing redshift. This is because both SN progenitor and galaxy ages are younger at higher redshifts compared to $z=0$ \citep{Rigault2013,Childress2014}: at low-redshift the mean \lq high mass\rq\ galaxy is old (or hosts old SNe) but at high-redshift the high mass hosts are younger (Fig.~\ref{fig:mass_age_z}). Thus, at high-redshift, hosts are younger at all stellar masses which should result in a reduced standardization  difference in the \citetalias{son_strong_2025} model \citep[see also][]{Rigault2018}. 

Fig.~\ref{fig:gamma1} shows the measured $\gamma$ in DES-SN5YR data as a function of redshift. This is measured from the difference between $\Delta\mu$ in low- and high-mass galaxies in each redshift bin without applying the $\gamma G_{{\rm host}}$ term (equation~\ref{eq:Ghost}). There is no evidence for any redshift evolution in $\gamma$.

The DES-SN5YR analyses fit these data via a simple, linear parametrization of
\begin{equation}
\gamma(z) = \gamma_0 + \gamma_1 z, 
\end{equation}
where $\gamma_0$ is the value of $\gamma$ at $z=0$ and $\gamma_1$ is a linear coefficient of any redshift evolution. DES-SN5YR find $\gamma_0=0.044 \pm 0.015$ and $\gamma_1=-0.028 \pm 0.034$ \citep{popovic_dark_2025}. We show the range of evolution allowed by the $\gamma(z)$ fit (i.e., the full uncertainty range on $\gamma_0$ and $\gamma_1$) in Fig.~\ref{fig:gamma1}. We find no significant evolution in $\gamma$ ($< 1 \sigma$) and thus no evidence that host galaxy stellar mass does not provide a sufficient standardization. Furthermore, when including the measured redshift evolution of $\gamma$, DES-SN5YR find a shift in the measurement of $w$ of $<0.01$\footnote{This shift is measured using SNe only to constrain $w$. The shift is smaller when combining with other probes.}.

To estimate the predicted $\gamma$ evolution from the alternative, age-bias model, we take the $\Delta\mu$--age relationship used by \citetalias{son_strong_2025} and apply it to SNe simulated with a simple galaxy evolution model following the method of \citet[][hereafter W22]{Wiseman2022}, outlined in more detail in Section~\ref{sec:age_review} and in Appendix~\ref{appendix:subsec:sn}. As expected, the \citetalias{son_strong_2025} model predicts a strongly evolving $\gamma$, inconsistent with that observed in the DES-SN5YR data. Most notably, the age-bias model predicts a low-redshift mass step of $\sim0.14$\,mag, $\sim6\,\sigma$ larger than that measured by DES5YR ($0.044\pm0.015$). The reason, as we show in the next section, is that the progenitors of SNe Ia observed in low-redshift surveys are not much older on average ($\sim$1--2\,Gyr) than the progenitors of SNe Ia in high-redshift surveys due to the shape of the SN Ia DTD.

\begin{figure*}
    \centering
    \includegraphics[width=0.48\textwidth]{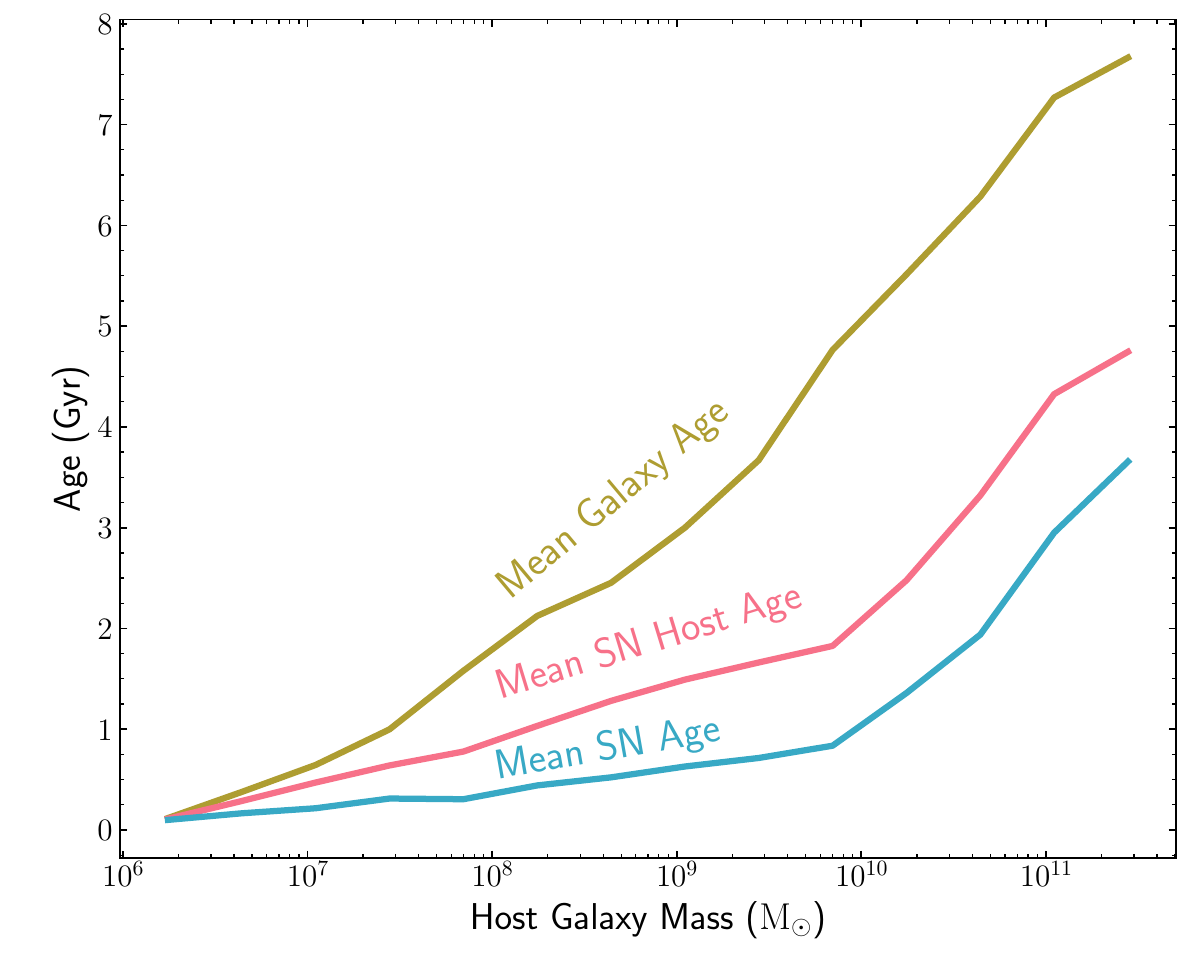}  \includegraphics[width=0.48\textwidth]{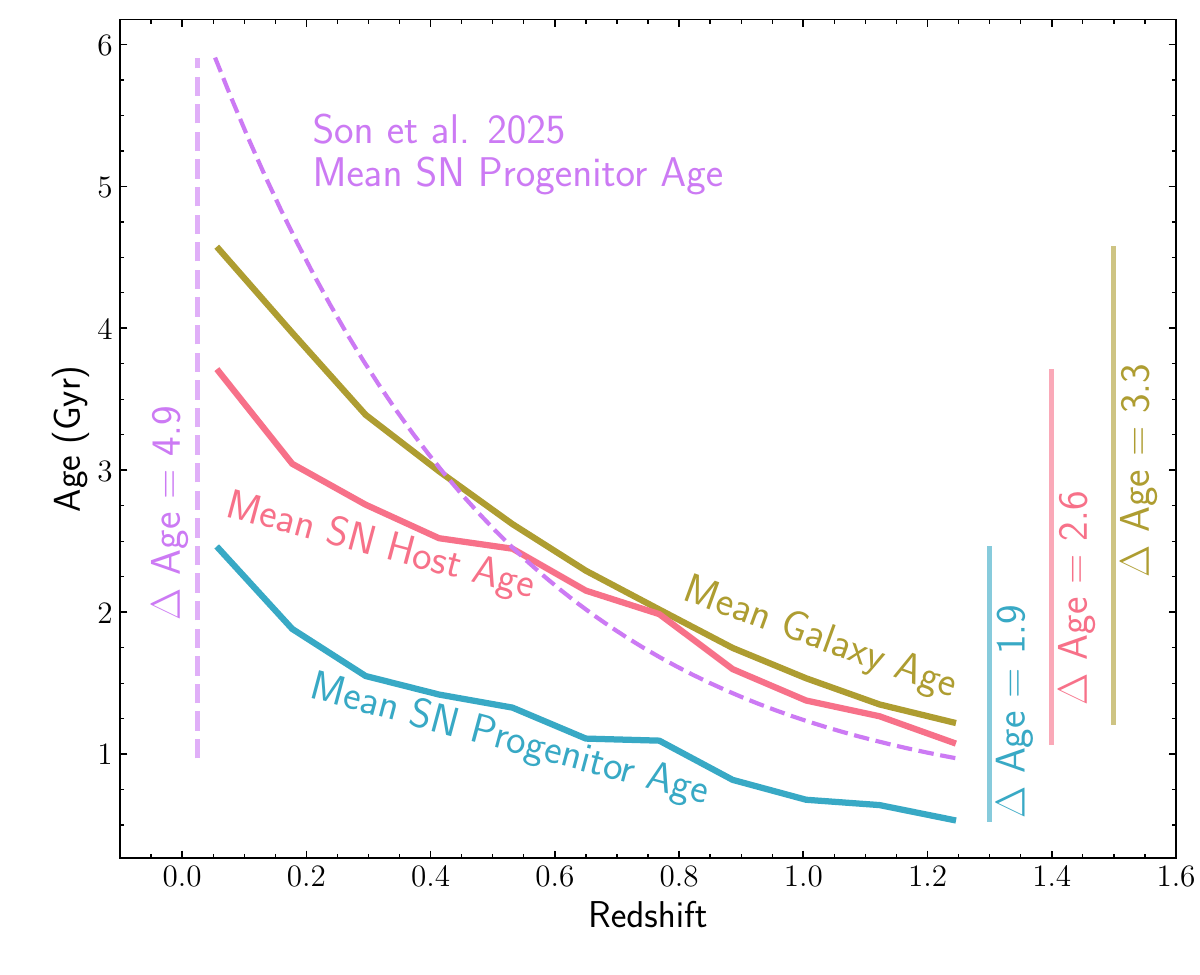}
\caption{Relationships between galaxy stellar masses, stellar and SN Ia progenitor ages, and redshifts, according to the physically-motivated models from \citetalias{Wiseman2022}. {\it Left}: Mean galaxy age (gold), mean SN Ia host galaxy age (pink), and mean SN Ia progenitor age (blue) as a function of galaxy stellar mass at $z=0$. The age of galaxies, as well as the SN Ia progenitors, is clearly correlated with the stellar mass;
{\it Right}: Mean galaxy age (gold), mean SN Ia host age (pink), and mean SN Ia progenitor age (blue) as a function of redshift. Due to the SN Ia delay time distribution (DTD), the SN age difference across the redshift range is half the galaxy age difference and 50 per cent smaller than the SN host age--redshift difference. In purple is the prediction from the DTD and cosmic SFH assumed by \citetalias{son_strong_2025}.
\label{fig:mass_age_z_single}}
\end{figure*}

\section{Modelling supernova--host-galaxy relationships as a function of redshift}
\label{sec:age_review}

In the previous section we showed how modern SN Ia cosmology analyses incorporate host galaxy environmental dependencies in SN Ia standardization. When these are included,  we find no empirical evidence for any signatures of the effects proposed by \citetalias{son_strong_2025}. In this section we use a complementary approach using a detailed, empirically calibrated model of the redshift evolution of SN Ia host galaxy properties to show what redshift-dependent changes in host age, SN progenitor age, or stellar mass, are allowed by observations. We show that these are far weaker than those suggested by \citetalias{son_strong_2025}.

\subsection{Modelling framework for SN Ia host galaxy masses and ages}

To demonstrate the relationships between redshift, SN Ia progenitors, and the ages and masses of the galaxies that host SNe Ia, we employ the simulations of \citet{Wiseman2022}. These in turn build upon the work of \citet{Childress2014} who presented an empirical model of the evolution of individual simulated galaxies and their stellar age distributions over cosmic time. We describe the models briefly in Appendix \ref{appendix:galaxy_sims}.

To calculate the SN Ia progenitor age distribution in a galaxy, the SFH of the galaxy is convolved with the SN Ia DTD, the function describing the probability of a given stellar population producing a SN Ia as a function of time since the stars were formed. The DTD is difficult both to predict from theoretical models and to measure, but can be inferred by measuring how the volumetric rate of SNe Ia changes with redshift \citep[e.g.,][]{Gal-Yam2004,Graur2011,Frohmaier2019} or how the specific SN rate varies among galaxies with different SFHs \citep{Totani2008, Graur2015, Wiseman2021,Castrillo2020}. Generally, these techniques find a DTD, $\Phi(t)$, consistent with a power-law of the form
\begin{equation}
    \Phi(t)=
    \begin{cases}
       0 &t<t_p\\
       A\,t^{\beta_\mathrm{DTD}} &t\geq t_p,\\
    \end{cases}
\end{equation}
where $t$ is the time since a stellar population formed, $A$ is a normalisation factor (the SN Ia \lq production efficiency\rq ), $\beta_\mathrm{DTD}$ the slope of the power-law, and $t_{\rm p}$ is the age of the stellar population before which SN Ia progenitor carbon-oxygen white dwarfs have not had sufficient time to form, and thus the probability of a SN Ia explosion is zero.

Observations suggest that $\beta_\mathrm{DTD}$ is around $-1$, and stellar evolution time scales set a lower limit for $t_{\rm p}$ of 30--40\,Myr. Such a power-law DTD has a key property that is important in this paper: it results in most SNe Ia originating from \lq young\rq\ ($<1$\,Gyr old) progenitors at all redshifts and, per unit stellar mass, galaxies with star formation therefore produce more SNe Ia than passive galaxies \citep[see][for relevant supporting observations]{oemler_type_1979,Mannucci2005,Sullivan2006}. This means that at any redshift and at any stellar mass, SN Ia host galaxy samples are dominated by galaxies with some star formation. However, the exact choice of the $\beta_\mathrm{DTD}$ and $t_{\rm p}$ parameters affect the distribution of SN Ia progenitor ages in a galaxy, which we discuss in further detail in Appendix~\ref{appendix:subsec:sn}.


\subsection{Issues with direct application of a redshift-dependent correction}

Our modelling of SN Ia and their host galaxies then reveals two key problems with the redshift-dependent standardization correction estimated by and used in \citetalias{son_strong_2025}. We discuss these in turn.

\subsubsection{Host galaxy age and SN Ia progenitor age are not interchangeable}

In the left panel of Fig.~\ref{fig:mass_age_z_single}, we show how our model galaxy ages, model SN host galaxy ages, and model SN progenitor ages correlate with host stellar mass at $z=0$. The relationship is strong as expected from data (see Fig.~\ref{fig:ages_and HRs}). The mean SN host galaxy age is younger than the mean galaxy age at all stellar masses, and that SN progenitor ages are younger still: estimates of galaxy age are not interchangeable with the SN progenitor age \citep[see also][]{Childress2014}. The effect of redshift on these relationships can be found in the Appendix (Fig.~\ref{fig:mass_age_z}).

This is because the age of a SN progenitor is not directly related to the age of the galaxy at the time of explosion due to the DTD: the DTD is dominated by short delay times, so SN samples are dominated by galaxies with some star formation. Galaxy age does not deterministically predict SN Ia progenitor age. (The only way in which this could be the case would be if the DTD was a delta function). \citetalias{son_strong_2025} therefore incorrectly assume that it is valid to apply their measured $\Delta\mu$--host-galaxy-age relationship to an estimate of redshift evolution of the SN progenitor ages. 

We also note the sharp increase in the slope of the galaxy and SN progenitor age correlations at $10^{10}\,{\rm M}_{\sun}$. This shows why, even if the environmental dependencies of SN luminosities were driven primarily by age, a mass standardization split at that mass effectively serves to standardize their luminosities.



\subsubsection{SN Ia progenitor ages show only modest redshift evolution}

In the right panel of Fig.~\ref{fig:mass_age_z_single}, we show the redshift evolution of the average mass-weighted age of the global galaxy population and of SN Ia host galaxies calculated from our models. Again, SN hosts are, on average, younger than the overall galaxy population and they also evolve less with redshift. The SN progenitor ages evolve even less than the SN host galaxy ages. SN progenitor ages, SN host ages, and galaxy ages have substantially different evolution with redshift and, again, should not be conflated. 

This means that the average SN Ia progenitor age difference between $z=0$ and $z=1$ predicted by \citetalias{son_strong_2025} is exaggerated at 5.3\,Gyr: the difference in SN Ia progenitor age in our models between $z=0$ and $z=1.2$ is only 1.9\,Gyr (Fig.~\ref{fig:mass_age_z_single}). This overestimation likely stems from the DTD employed by \citetalias{son_strong_2025}, with a prompt timescale of $t_{p}=0.3$\,Gyr and a power-law slope of $\beta_{\rm DTD}=-1.0$, as per \citet{Childress2014}, rather than 30--40\,Myr prompt time set primarily by stellar and binary evolution \citep{hachisu_delay-time_2008,hachisu_young_2008}, and steeper value ($\sim-1.13$) of the power law slope indicated by recent DTD measurements. Appendix~\ref{appendix:subsec:sn} discusses the choices of DTD parameters in more detail.


\subsection{Why a host-age driven effect predicts a redshift-evolving mass step}

The mass step is calculated at a fixed stellar mass ($P_\mathrm{step}$ in equation~\ref{eq:Ghost}), but because of the correlations between galaxy age, redshift, and stellar mass, the mean age of galaxies on either side of that threshold changes as a function of redshift. At low redshift, there is a large difference in mean age between low- and high-mass galaxies, but at high redshift this difference is smaller (Fig.~\ref{fig:step_frac_evol}). The \citetalias{son_strong_2025} hypothesis, that galaxy age drives the mass step, results in a strong mass step in low-redshift samples because of the strong galaxy age--mass correlation and the large difference between the ages of the low- and high-mass galaxies. However, at high redshift, because the host ages are  more uniform regardless of their stellar mass, there is minimal standardization difference between the SNe in low- and high-mass galaxies, resulting in no mass step. In Appendix~\ref{appendix:mass_frac} we show how this effect is produced as an outcome of the overall change in the age of SN hosts as a function of redshift.

To predict the scale of redshift evolution of the mass step predicted by the \citetalias{son_strong_2025} hypothesis (the line in Fig.~\ref{fig:gamma1}), we draw SNe from galaxies in our simulation following the SFH and DTD.
We then ascribe a `luminosity offset' based on the mass-weighted age of the host according to the $0.03$\,mag\,Gyr$^{-1}$ slope used by \citetalias{son_strong_2025}. As a function of redshift, we then measure the average offset in low and high mass galaxies (i.e., the mass step), which we plot against the measured redshift evolution.  As is apparent, a strong evolution in the mass-step expected by a strong change in SN progenitor age is not seen in the DES-SN5YR data.

\subsection{Fraction of SNe in high mass host galaxies}

\begin{figure}
    \centering
    
    \includegraphics[width=0.499\textwidth]{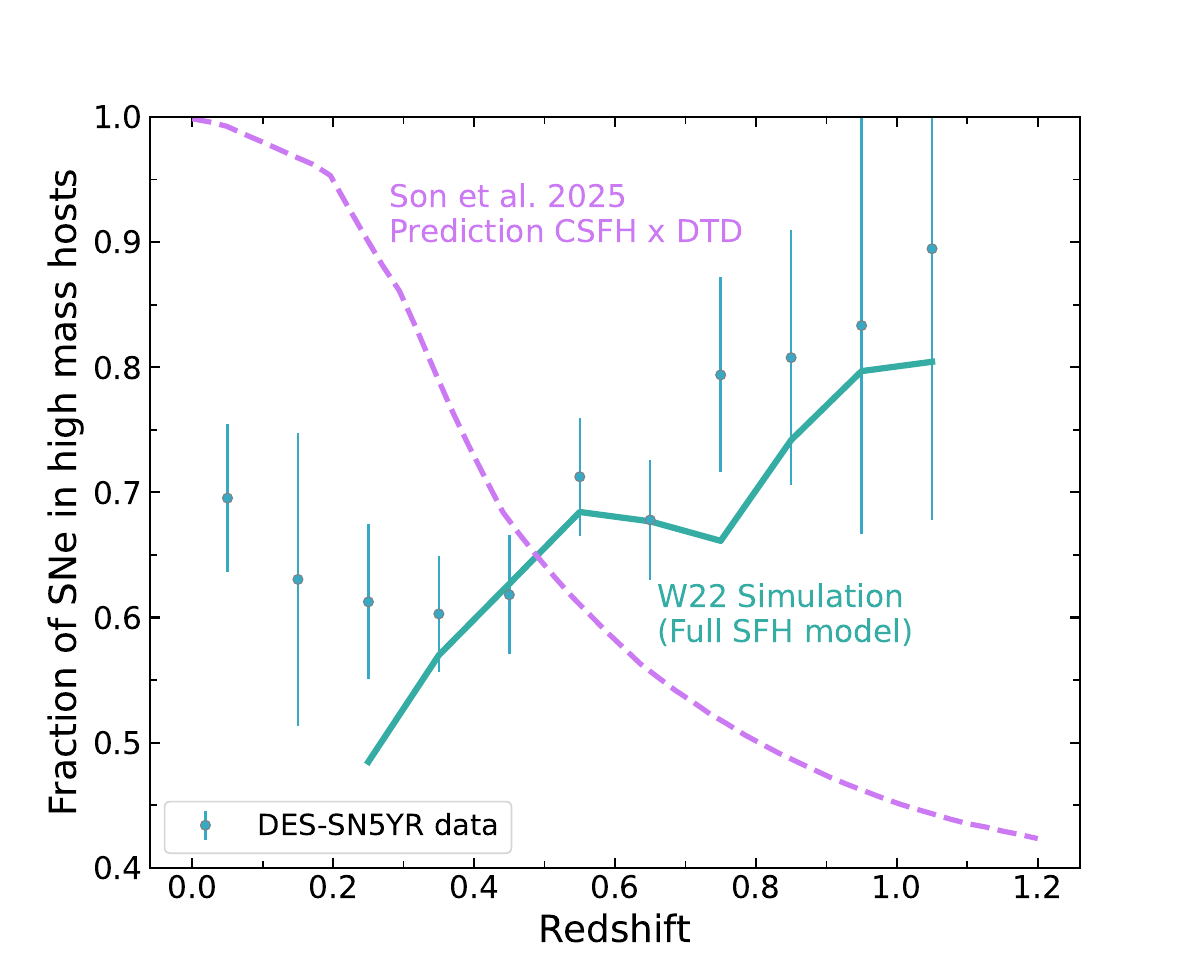}
\caption{Evolution of the fraction of SNe Ia in high stellar mass galaxies as a function of redshift.  Data are taken from the full DES-SN5YR dataset, including low-redshift surveys. The solid line is the prediction from a simulation with full treatment of galaxy quenching, starbursts, dust, and SN and galaxy selection effects from \citetalias{Wiseman2022}. The dashed line is a prediction based on estimating how the fraction of SNe Ia in high mass galaxies would evolve with redshift given the large difference in SN Ia ages predicted by the DTD and cosmic SFH model used by \citetalias{son_strong_2025}.
\label{fig:high_mass_frac}}
\end{figure}

A full explanation of the lack of redshift evolution of Hubble residuals and the mass step in observations, compared to that claimed by \citetalias{son_strong_2025}, is a combination of the effects described thus far in this section: host galaxy and SN Ia progenitor ages are not interchangeable, and host galaxy and SN Ia progenitor ages do not evolve as strongly with redshift as claimed. A final check of the \citetalias{son_strong_2025} prediction versus the data can be made by considering the fraction of SNe Ia that are observed in high-mass hosts as a function of redshift.

If SN Ia host ages vary as strongly with redshift as predicted by \citetalias{son_strong_2025} then the fraction occurring in low and high-mass hosts will also vary with redshift. The evolution of this high-mass fraction is directly predictable by the \citetalias{Wiseman2022} model as individual galaxies are treated independently; however, using the DTD convolved with cosmic SFH model as in \citetalias{son_strong_2025} does not allow for the comparison of host stellar mass, since the cosmic SFH integrates over stellar mass. To estimate the evolution of the high-mass fraction in the DTD$\times$cosmic-SFH model, we take galaxies in the \citetalias{Wiseman2022} model at a fixed redshift $z=0.5$ and compute the fraction of high-mass galaxies as a function of SN Ia progenitor age. We then take the predicted SN Ia ages as a function of redshift from the \citetalias{son_strong_2025} DTD$\times$cosmic-SFH model and interpolate from the \citetalias{Wiseman2022} SN Ia age distribution to estimate a high mass fraction.

The fraction of SNe in galaxies above the mass-step threshold $\log(M_*/{\rm M}_{\sun})=10$ is shown in Fig.~\ref{fig:high_mass_frac}, again using the DES-SN5YR dataset. The \citetalias{Wiseman2022} model, which takes into account stochastic galaxy processes and survey selection effects, reproduces the evolution of the high-mass fraction well from $z>0.3$ with no tuning. This simulation was designed to reproduce the DES-SN5YR data so does not extend to low redshifts where survey selection effects are different.

The estimated evolution of the high-mass fraction using the \citetalias{son_strong_2025} DTD and cosmic SFH with no selection effects is a very different shape to the observations. We have tested how this prediction is affected by the choice of redshift with which to estimate the SN age--mass-fraction function, and the overall trend is always the same. Although this figure does not directly explain the lack of evolution in the mass step in observations, it exemplifies the nuance and care required in predicting the evolution of SN host galaxy properties, in particular the treatment of individual galaxies rather than using universal functions, and the importance of including survey selection effects.

\subsection{Correlation versus causation}
The assumption underpinning the claim of \citetalias{son_strong_2025} is that the host galaxy age {\it drives} the luminosity variation of SNe Ia. While the above analysis shows that other assumptions they make exaggerate the strength of any redshift evolution, one would expect no such evolution at all if the observed Hubble-residual--age relationship is driven by an underlying effect that evolves less, or not at all, with redshift. There is a subtle but critical difference between the causation and correlation of Hubble residual relationships: Hubble residuals before mass-standardization correlate with host galaxy age (Fig.~\ref{fig:ages_and HRs}), but they may not be caused by it; Hubble residuals correlate with many other host galaxy parameters.

The physics of the explosions, or the way we observe them, are not directly influenced by the age of the galaxy (or, for that matter, its stellar mass). Rather, there are unknown parameters, which correlate with the host galaxy age and the SN physics, or the way we observe the SNe. The true driving cause of the relationship is uncertain, and if it does not evolve with redshift or evolves less with redshift than host galaxy age, then applying a correction based on the host age is incorrect.

There may also be other systematic effects. For example, possible deficiencies in the linear standardization of SNe Ia from equation~\ref{eq:simple_tripp} are beginning to emerge with new high-fidelity low-redshift datasets, particularly non-linearities in the $\alpha x_1$ term in low-$x_1$ SNe Ia \citep{garnavich_connecting_2023,larison_environmental_2024,newsome_mapping_2024,ginolin_ztf_2025-1}. Such low-$x_1$ SNe Ia preferentially occur in massive and passive host galaxies \citep{Hamuy2000,Sullivan2010}, reinforcing the incorrectness in assuming that any effects observed in such SN Ia populations can be attributed to age alone.

\section{Conclusions}\label{sec:conclusion}

We have tested whether the correlation between SN Ia Hubble residuals and host galaxy ages claimed by \citetalias{son_strong_2025} can induce an apparent redshift evolution in SN Ia standardization that biases cosmological inferences. While \citet{lee_evidence_2022} and \citetalias{son_strong_2025} suggested that such an effect could remove the need for dark energy or otherwise significantly modify cosmological inferences, our analysis shows that these effects are both already accounted for, and exaggerated. We find:

\begin{enumerate}
    \item Applying mass-standardization and bias-corrections in line with state-of-the-art cosmological analyses reduces the strength of the Hubble-residual--age relationship and renders it insignificant;
    \item An effect driven by galaxy age predicts a redshift evolution in the size of the mass step; we show that measurements of the evolving mass step are inconsistent with the \citetalias{son_strong_2025} prediction at $\sim6\,\sigma$;
    \item We identify a number of inaccurate assumptions present in the \citetalias{son_strong_2025} analysis, namely the direct application of a galaxy-age measurement to an SN-age redshift evolution, and an overestimation of the redshift evolution of SN ages due to assumptions about the SN Ia delay-time distribution;
    \item We reiterate the need for careful modelling of galaxy star-formation histories and survey selection effects when predicting the evolution of SN Ia luminosities and host galaxy parameters.
\end{enumerate}

SN Ia cosmology is a mature field in which great care is taken to address complex and often hidden systematic effects. Testing for the robustness of the methods is a critical undertaking, but one that requires equal measures of care. We look forward to the future of SN Ia cosmology as we enter the era of extreme datasets from the Zwicky Transient Facility \citep{Bellm2019,rigault_ztf_2025}, the Vera C. Rubin Observatory \citep{ivezic_lsst_2019,lochner_impact_2022,gris_designing_2023}, the TiDES survey on the 4-metre Multi-Object Spectroscopic Telescope \citep{frohmaier_tides_2025}, and the Nancy Grace Roman Space Telescope \citep{kessler_cosmology_2025}, where such care will be of more importance than ever before.

\section*{Acknowledgements}
We thank the referee for their constructive comments. We are also grateful to Chris Lidman, Matt Middleton and Saul Perlmutter for insightful comments.
P.W. is grateful for the support from the Science and Technology Facilities Council (STFC) grant ST/Z510269/1 and MSu from grant ST/Y001850/1. TMD acknowledges support from the Australian Research Council through the Centre of Excellence for Gravitational Wave Discovery (OzGrav), project number CE230100016. L.G. acknowledges financial support from CSIC, MCIN and AEI 10.13039/501100011033 under projects PID2023-151307NB-I00, PIE 20215AT016, and CEX2020-001058-M. IH gratefully acknowledges support from the STFC grant ST/Y001230/1. This project has received funding from the European Union’s Horizon Europe research and innovation programme under the Marie Skłodowska-Curie grant agreement No 101205780. LK acknowledges support for an Early Career Fellowship from the Leverhulme Trust through grant ECF-2024-054 and the Isaac Newton Trust through grant 24.08(w). Supernova research at Rutgers University is supported in part by NSF grant AST-2407567 and DOE award DE-SC0010008. SWJ also gratefully acknowledges support from a Guggenheim Fellowship.

\section*{Data Availability}

No new original data are presented in this paper, and all data used are publicly available in the references given.



\bibliographystyle{mnras}
\bibliography{PWbib_offline} 




\appendix

\section{Simulations}\label{appendix:galaxy_sims}
We base our simulations on the framework developed in \citet{Wiseman2021} and \citet{Wiseman2022}, itself derived from the concepts laid out in \citet{Childress2014}: see these papers for a full description. Here we provide a brief overview of the simulations, which are produced in several stages:

\begin{enumerate}
    \item a semi-empirical model of galaxies evolving with cosmic time;
    \item a reconstruction of galaxy observables from star-formation histories;
    \item the generation of SNe Ia within galaxies according to the SFH and DTD allowing for the tracing of SN Ia progenitor and galaxy ages;
    \item the generation of SN Ia light-curve parameters according to the SN progenitor and galaxy properties.
\end{enumerate}

\subsection{Galaxies}\label{appendix:subsec:gals}
Galaxy simulations are described in full in \citet{Wiseman2021}, and are based on the original prescription in \citet{Childress2014}. Each galaxy is seeded at $10^6\,{\rm M}_{\sun}$ at some formation time and grows according to empirical relationships between star-formation rate, stellar mass, and redshift. A basic quenching \lq penalty\rq\ is applied such that star-formation shuts off as galaxies grow more massive. Stellar mass loss to SNe (of all types) and compact objects is accounted for. This model was updated in \citet{Wiseman2022} to include a stochastic prescription for quenching, and is able to reproduce both the overall cosmic SFH \citep[][]{behroozi_average_2013, Madau2014}, the star-forming main sequence, and a population of passive galaxies. For each galaxy, the stellar age distribution is known with a granularity of 0.5\,Myr. These SFHs are combined with stellar population synthesis codes to provide estimates for galaxy spectra and global galaxy broadband colours, and reproduce the bimodal distribution of red and blue galaxies.
The simulations reproduce the overall mass assembly of galaxies with cosmic time.

\subsection{Type Ia supernovae}
\label{appendix:subsec:sn}

To choose galaxies in the simulation to become SN Ia hosts, the relative rate of SNe Ia in each host is derived from the sum of the convolution of the SFH and the SN Ia DTD (i.e., integrating the SN progenitor age distribution with respect to age). We use the $\beta_{\rm DTD}=-1.13$ and $t_p=0.04$\,Gyr. The total number at a given stellar mass is this summed convolution multiplied by the stellar mass function, for which we use the redshift-dependent ZFOURGE functions \citep{Tomczak2014}. This step constructs the true {\it intrinsic} SN Ia host distribution. The progenitor age of each of these SNe is determined by drawing at random from the SN Ia progenitor age distribution.

The choice of the parameters defining the DTD can have a significant impact on the details of the simulations. Recent measurements using a variety of complementary methods tend to $\beta_{\rm DTD}<-1$, \citep{Heringer2019, Castrillo2020, Freundlich2021,  Wiseman2021, Chen2021}. The prompt time is challenging to measure, and is expected to lie between $0.04-0.3$\,Gyr with most measurements preferring the lower end of that range: \citet{aubourg_evidence_2008} found an upper limit of 180\,Myr; \citet{Wiseman2021} found $t_{\rm p}$ to be consistent with 0.04\,Gyr;  \citet{Chen2021} found $t_p=0.12^{+0.14}_{-0.08}$\,Gyr; \citet{Castrillo2020} found $t_p=0.05^{+0.10}_{-0.03}$\,Gyr. \citet{Wiseman2021} used the canonical 40\,Myr in subsequent simulations that accurately reproduce SN Ia host galaxy populations \citep{Wiseman2022}.

\begin{figure}
    \centering
    \includegraphics[width=0.48\textwidth]{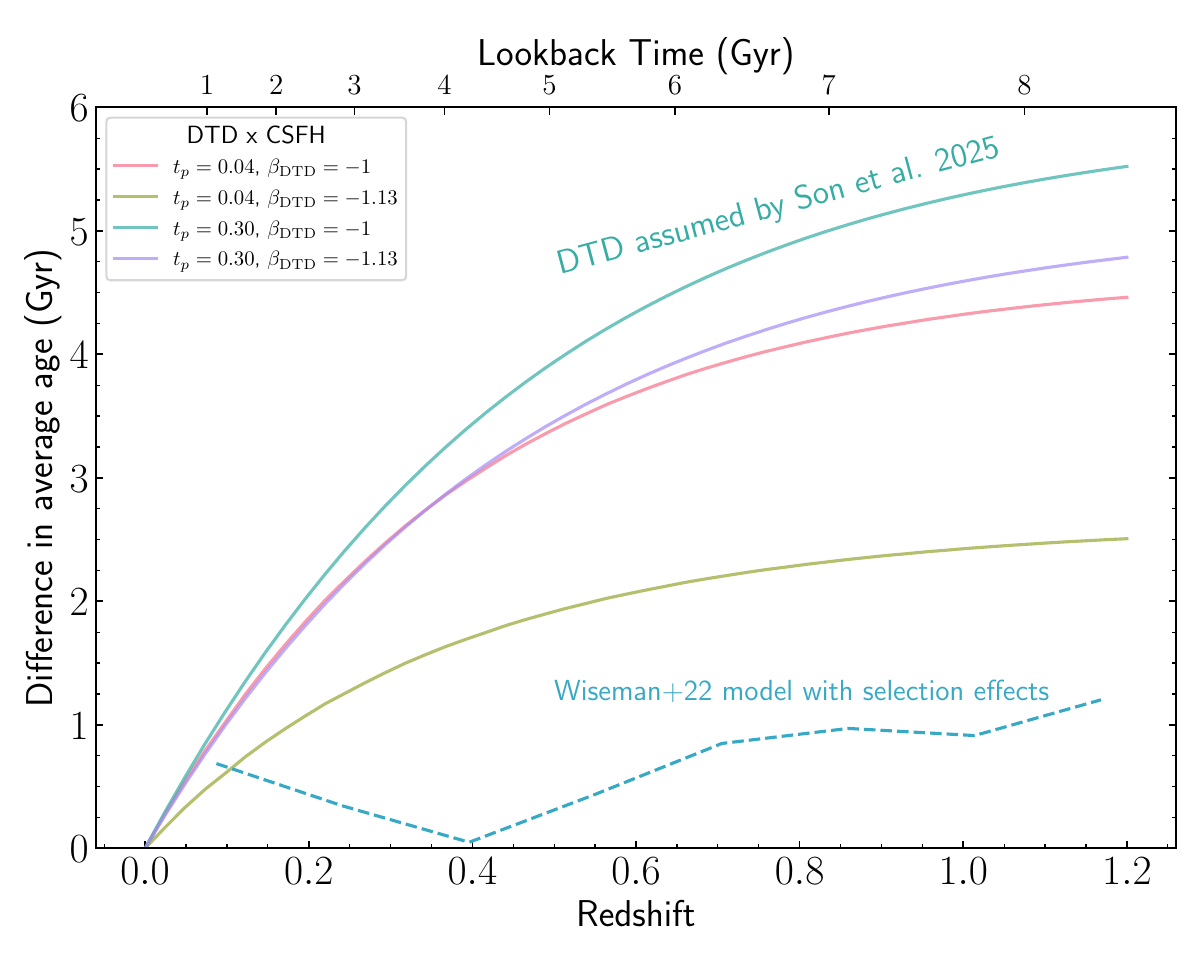}
\caption{Differences in the median SN Ia progenitor age as a function of redshift (lookback time) under different DTD assumptions.  With $\beta_{\rm DTD}=-1$, $t_p=300$\,Myr the difference in median progenitor age (or median delay time) between SN populations at $z=0$ and $z=1$ is 4.5\,Gyr. In contrast, with $\beta_{\rm DTD}=-1.13$, $t_p=40$\,Myr the difference is closer to 1.5\,Gyr. The dashed line is the median progenitor age of SNe in a simulation with full treatment of galaxy quenching, starbursts, dust, and SN and galaxy survey selection effects.
\label{fig:dtd_evol}}
\end{figure}

Fig.~\ref{fig:dtd_evol} illustrates the difference between the evolution in average age given the DTD parameters assumed by \citetalias{son_strong_2025} and those inferred by \citet{Wiseman2021}. Each line has been produced by convolving a \citet{behroozi_average_2013} cosmic SFH (as adopted by \citetalias{son_strong_2025}) with a DTD, and implies that the magnitude of any putative standardization evolution with redshift is highly sensitive to the adopted DTD. Overestimating the prompt timescale and underestimating the strength of the power-law slope exaggerates the expected age difference across redshift, over-predicts the number of low-redshift SNe Ia in quiescent galaxies, and exaggerates the inferred impact on cosmology of an age-dependent bias. We refer back to the right-hand panel of Fig.~\ref{fig:mass_age_z_single} where we have plotted the progenitor age evolution predicted by \citetalias{son_strong_2025}, which is clearly overestimated by several Gyr compared to our detailed modelling, to the extent that \citetalias{son_strong_2025} predict that SN progenitors evolve more than the average age of all galaxies in our model.

The use of the cosmic SFH to approximate the evolution of galaxy ages with cosmic time is also problematic. While in principle convolving the DTD with the cosmic SFH should result in the same SN progenitor age distribution as producing individual SFHs and convolving each with the DTD and the galaxy stellar mass function, in practice, SN surveys are subject to selection effects. Individual galaxies experience quenching and starbursts, are obscured by dust, and require identification and measurements in order to be included in SN host samples. The dashed line in Fig.~\ref{fig:dtd_evol} shows the recovered SN progenitor age evolution in a simulation that includes realistic modelling of evolutionary processes as well as selection effects. The difference in median progenitor age observed between low and high redshift is minimal ($\sim$1\,Gyr).


\subsection{Age and mass fraction evolution with redshift}
\label{appendix:mass_frac}
Here we show how the redshift evolution of average SN Ia host age naturally predicts an evolving stellar mass step, if the step were to be entirely driven by the host age. 

\begin{figure}
    \centering
    \includegraphics[width=0.49\textwidth]{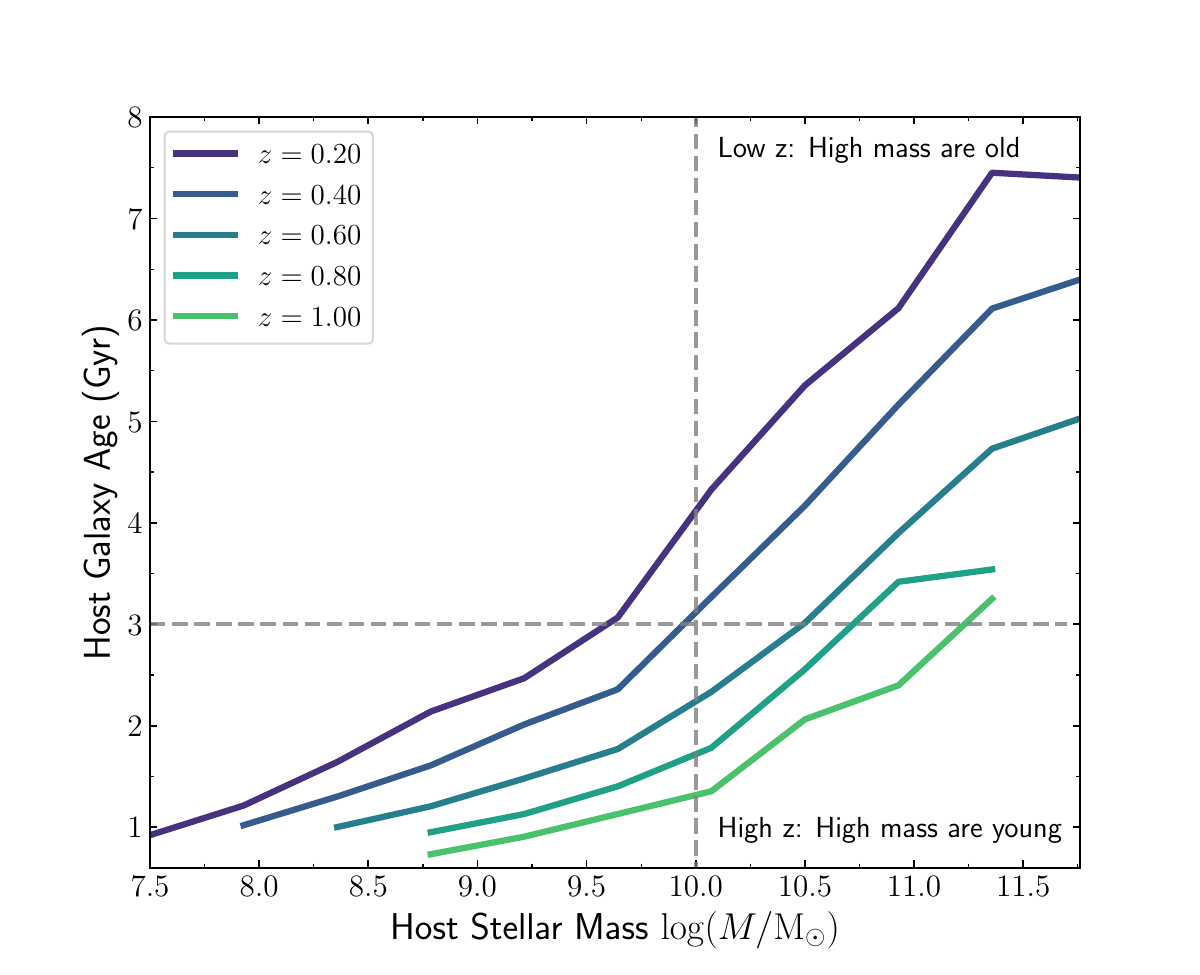}
\caption{SN Ia host galaxy ages as a function of their stellar mass for different redshifts. At fixed stellar mass, hosts are younger at higher redshift.
\label{fig:mass_age_z}}
\end{figure}

\begin{figure}
    \centering
    \includegraphics[width=0.49\textwidth]{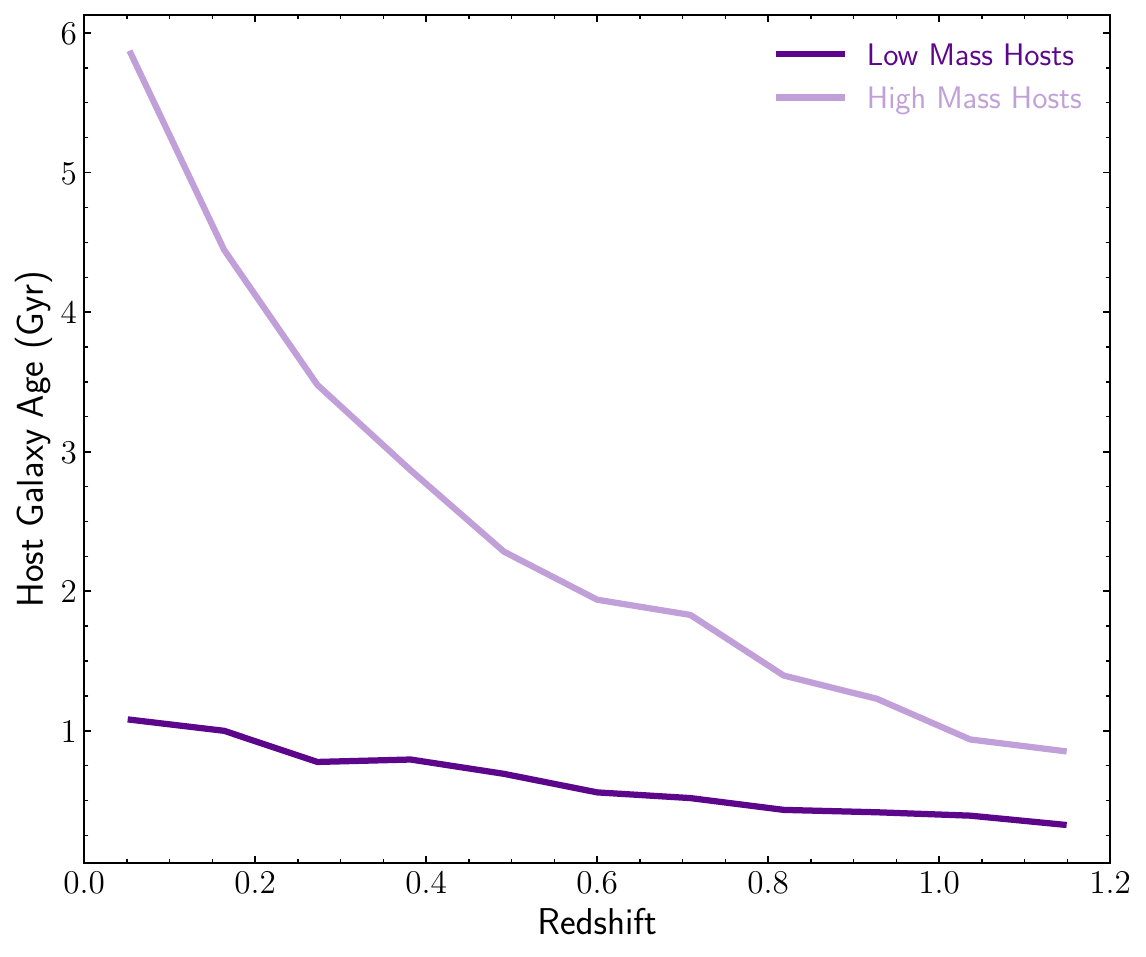}
\caption{The average age of SN Ia host galaxies on either side of the mass step ($10^{10}\,{\rm M}_{\sun}$), as a function of redshift. At high redshift, the average high-mass galaxy is still relatively young. Thus, if the SN Ia luminosity differences are caused by gradients in luminosity across a large range of galaxy ages, high-redshift hosts will show very little difference between the low and high-mass bins, resulting in a smaller mass step.
\label{fig:step_frac_evol}}
\end{figure}

Fig.~\ref{fig:mass_age_z} is similar to Fig.~\ref{fig:mass_age_z_single} except we now show the average SN Ia host age at five different redshifts. At low redshift, low-mass galaxies are young while high-mass galaxies are old. At high redshift, even the most massive galaxies are comparatively young. The fraction of galaxies either side of the canonical $10^{10}\,{\rm M}_{\sun}$ host mass split is shown in Fig.~\ref{fig:step_frac_evol}. At low redshift, even if the step is driven by host age, the old galaxies are massive and young galaxies are low mass, meaning the step will be evident when traced by mass. At high redshift, since all the high mass galaxies are also young, there would be little luminosity difference between SNe Ia in low and high-mass hosts. 

\bsp	
\label{lastpage}
\end{document}